\documentclass[preprint,12pt]{aastex}
\usepackage[]{graphicx}
\shorttitle{Formation of GEMS in superbubbles}
\shortauthors{Westphal \& Bradley}

\begin{document}
\newif\ifincludefigures
\newif\ifnofigures

\includefigurestrue
\includefiguresfalse




\title{Formation of GEMS from shock-accelerated crystalline dust in superbubbles}

\author{A. J. Westphal}
\affil{Space Sciences Laboratory, University of California,
    Berkeley, CA 94720}
\email{westphal@ssl.berkeley.edu}
    
    \and

\author{J. P. Bradley }
\affil{Institute for Geophysics and Planetary Physics, Lawrence Livermore National Laboratory, Livermore, CA 94035}
\email{jbradley@igpp.ucllnl.org}

\begin{abstract}
Interplanetary dust particles (IDPs) contain enigmatic sub-micron components
called GEMS (Glass with Embedded Metal and Sulfides).  The compositions and structures of GEMS   indicate that they have been processed by exposure to ionizing radiation but details of the actual irradiation environment(s) have remained elusive.  Here we propose a mechanism and astrophysical site for GEMS formation that explains for the first time the following key properties of GEMS; they are stoichiometrically enriched in oxygen and systematically depleted in S, Mg, Ca and Fe (relative to solar abundances), most have normal (solar) oxygen isotopic compositions, they exhibit a strikingly narrow size distribution (0.1-0.5 $\mu$m diameter), and some of them contain ``relict''
crystals within their silicate glass matrices.  We show that the compositions, size distribution, and survival of relict crystals are inconsistent with amorphization by particles accelerated by diffusive shock acceleration.  Instead, we propose that GEMS are formed from crystalline grains that condense in stellar outflows from
massive stars in OB associations,  are accelerated in encounters with frequent supernova shocks inside the associated superbubble, and are implanted with atoms from the hot gas in the SB interior.  We thus reverse the usual roles of target and projectile.  Rather than being bombarded at rest by energetic ions, grains are accelerated and bombarded
 by a nearly monovelocity beam of atoms as viewed in their rest frame.
Meyer, Drury and Ellison have proposed that galactic cosmic rays originate from ions
sputtered from such accelerated dust grains.  
We suggest that GEMS are surviving members of a population of fast grains that constitute
the long-sought source material for galactic cosmic rays.    Thus, representatives of
the GCR source material may have been awaiting discovery in cosmic dust labs for the last thirty years.
\end{abstract}


\section{GEMS:   enigmatic grains in IDPs}
%
%
		
GEMS consist of amorphous silicate
glass grains with non-stoichiometric concentrations of oxygen, and depletions of Mg, S, Ca, and Fe, relative to Si, compared to solar system abundances.  They also contain nanometer-scale
inclusions of kamacite (Fe-Ni metal) and pyrrhotite ($\sim$FeS with up to 2\% Ni).   
GEMS are found in a strikingly narrow size range 
(100-500nm); specifically, small GEMS ($<100$nm) are conspicuously absent from IDPs.  
GEMS are often pseudo-euhedral despite the fact that 
they are structurally amorphous (Fig. 1). This property suggests that GEMS were originally 
individual crystalline mineral grains that were amorphized by a large fluence of ionizing 
radiation before they were incorporated into the solar nebula.
Some GEMS contain  internal relict grains that mimic the external shape and orientation
of the euhedral structure of the entire grain.  In other words, these GEMS are pseudomorphs (Fig. 1).

 \begin{figure}[h!]
\begin{center}
\rotatebox{0}{\resizebox{!}{8cm}{\includegraphics*{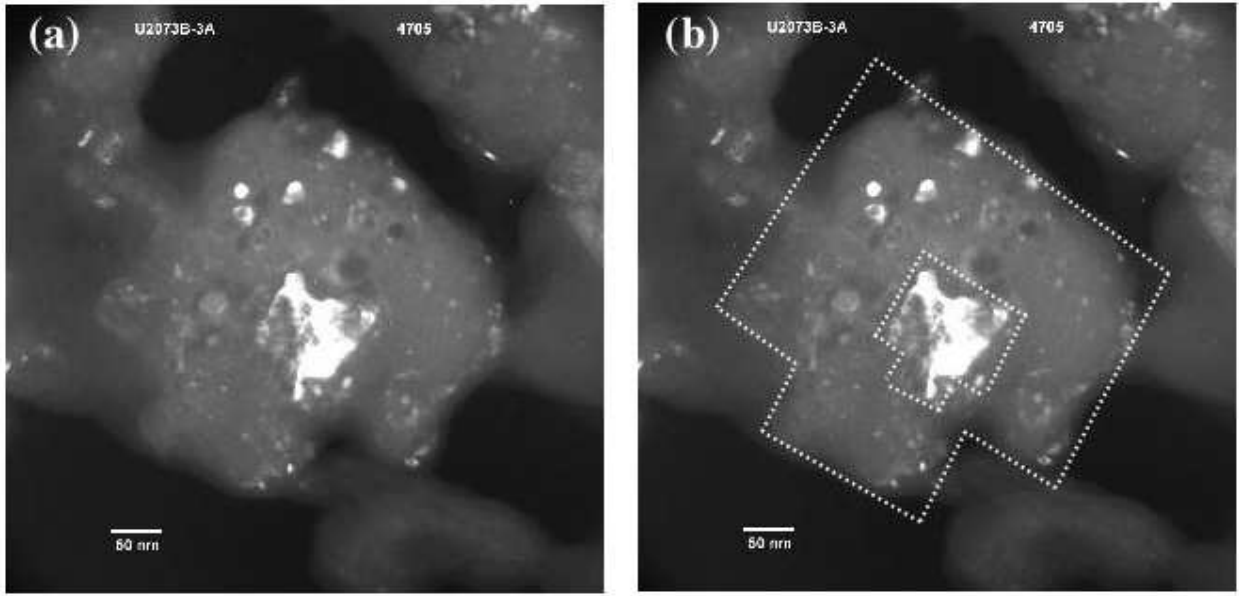}}}
\caption{(a) Darkfield micrograph of GEMS with an embedded relict pyrrhotite (FeS) crystal.Ê (b) Same GEMS as in (a) with dotted lines highlighting the approximate outlines of the relict pyrrhotite crystal and the GEMS itself.Ê Comparison of the lines indicates that the GEMS is a pseudomorph of the relict pyrrhotite crystal.  (See also \protect\citet{bradley-dai04} and \protect\citet{bradley99}.)} \label{gemsimage}
\end{center}
\end{figure}

Here we propose a mechanism for the formation of GEMS from shock-accelerated mineral grains
in the hot, low-density ISM.  The organization of the paper is as follows.
In section \ref{bombardment} we review the phenomenology of atomic bombardment of submicron grains.
In section \ref{isolated-supernova} we consider and reject a model in which GEMS are produced
by encounters with isolated supernova shocks.
In section \ref{shock-acceleration} we review the theory of dust acceleration by astrophysical
shocks.   In section \ref{superbubbles} we propose an astrophysical site --- the hot, low-density cavities
blown by OB associations,  called superbubbles --- in which GEMS are synthesized.
In section \ref{connection} we discuss the connection between GEMS and galactic cosmic rays.
Finally, in section \ref{discussion} we conclude, and suggest new measurements that would test our
hypothesis.

 \section{Grain modification by high-velocity atomic bombardment}
 \label{bombardment} 

The fact that many GEMS exhibit a pseudo-euhedral shape but are mostly amorphous or exhibit amorphous rims
 implies that GEMS are single crystals that have been at least partially amorphized by intense irradiation (e.g. Fig. 1)\citep{bradley94}.  
In an effort to make a laboratory simulation of  the amorphization of dust grains in the ISM,  \citet{demyk} and \citet{carrez} have  demonstrated
 that olivine can be efficiently amorphized by 1-12.5 keV/amu (400-1500 km sec$^{-1}$) He ion bombardment with fluences $\ge 10^{16}$ cm$^{-2}$.   A natural candidate for the ionizing
 radiation that could produce amorphization is the population of relativistic ions 
 that fill interstellar space --- the galactic cosmic rays (GCRs).  Because low-energy GCRs are
 excluded from the heliosphere,  the GCR flux has only
  been measured for energies $> $100 MeV/amu. However, the GCR fluence over the lifetime of dust
  in the ISM is probably insufficient to  produce the observed amorphization\citep{jones2000}.

\subsection{Production of pseudomorphs and survival of relict crystals}

Relict crystals have been observed in 10-20\% of GEMS \citep{bradley-dai04}
 We point out that the survival of relict { internal crystals} is also inconsistent with 
 amorphization by GCRs.   The sharp boundaries of
  relict { crystals} require a very sharp cutoff in bombarding ion energy, so that the grains
  are amorphized only to a depth corresponding to the maximum depth of penetration (range) of the bombarding ions,
  and no further.  GCRs with energies up to 
 at least $\sim10^{15}$eV are accelerated by diffusive shock acceleration in supernova  (SN) shocks (see,
 e.g., \citet{gaisser}.)  This mechanism inevitably 
 produces a very smooth, power-law spectrum, with no sharp cutoff in flux with increasing energy.  
  
 \begin{figure}[h!]
\begin{center}
\rotatebox{0}{\resizebox{!}{10cm}{\includegraphics*{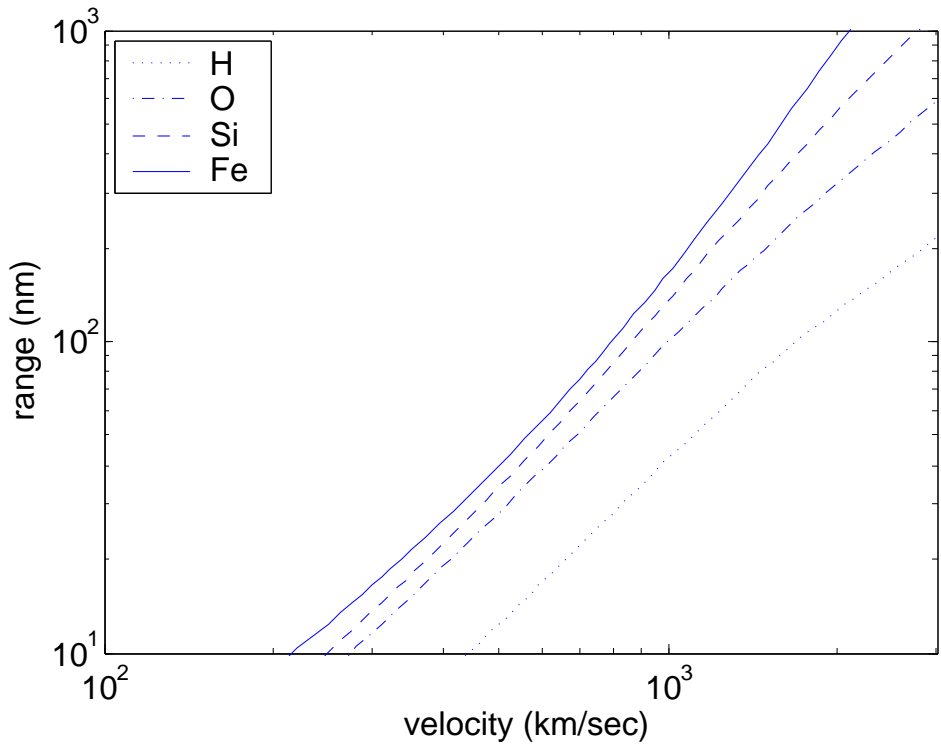}}}
\caption{Range in nm of ions H, O, Si and Fe atoms in troilite as a function of ion velocity. 
 Ranges were calculated using SRIM 2003.26\protect\citep{srim}.} \label{range-velocity}
\end{center}
\end{figure}

The amorphous rims in GEMS vary continuously in thickness
 from $\sim$10nm  
 to the radius of the entire grain, in which case no relict { crystal} survives.
  There may be an observational bias:  a GEMS with a 10nm 
  rim could be mistaken for an ordinary unprocessed crystalline mineral grain.  However, completely 
 amorphized GEMS are unambiguous even in the absence of a relict { crystal}.    In Fig. \ref{range-velocity} we show the range  of
cosmochemically abundant ions as a function of ion velocity in FeS (density = 4.6 g cm$^{-3}$).    At constant velocity, Fe has the largest range of any
significantly abundant element.   A range of 100nm for Fe atoms  in troilite corresponds 
approximately  to a velocity of $\sim$ 800 km sec$^{-1}$.

Here we suggest that GEMS are formed from  euhedral crystals  that are exposed
to fast ($\sim$ 1000 km sec$^{-1}$) atoms which have a single-valued velocity at any given time.   
In section \ref{superbubbles} we discuss the astrophysical site in which this bombardment can occur naturally.

\subsection{Modification of bulk  grain chemistry due to monovelocity bombardment by SN ejecta}

\begin{figure}[h!]
\begin{center}
\resizebox{!}{10cm}{\includegraphics*{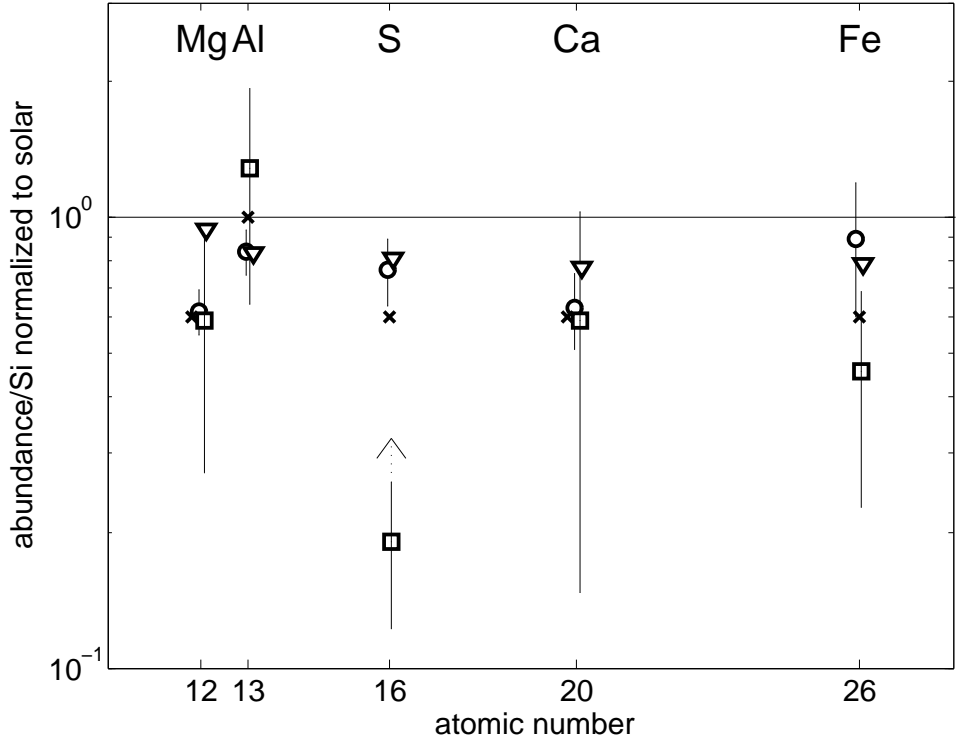}}  
\caption{Comparison of major element abundances, all with respect to Si, normalized to solar values,
in GEMS (squares and crosses) and in IMF-averaged core-collapse SN ejecta (circles). Abundances in GEMS (squares) were measured by \protect\citet{bradley-ireland}.
The S abundance is a lower limit because S is volatile and some GEMS contain secondary magnetite (Fe$_3$O$_4$) formed by partial oxidation of FeS (and loss of S) due to frictional heating during atmospheric entry.   Measurements as reported by \protect\citet{keller-messenger} (without errors) are shown as crosses.  Core-collapse SN yields (13-35 M$_{\odot}$) from Limongi and Chieffi were integrated over the Salpeter IMF (circles).  Error bars in SN yields were calculated assuming that the uncertainty in
yield for a SN progenitor mass was equal to the dispersion among the models for that mass.  Bulk
measurements for GEMS-rich IDPs\protect\citep{schramm} are also shown for comparison (triangles).} \label{LandC}
\end{center}
\end{figure}

Although GEMS-rich chondritic porous (CP) IDPs are believed to be the most chemically and isotopically primitive (unaltered) meteoritic materials, they exhibit systematic deviations from solar abundances.  In a study of the bulk compositions of 90 CP IDPs Schramm {\em et al.} found that they are stoichiometrically enriched in O and systematically depleted in Mg, S, Ca and Fe \citep{schramm}.  These 
compositional trends are even more pronounced in the individual GEMS within CP IDPs.
\citet{bradley94}, \citet{bradley-ireland} and \citet{keller-messenger} have measured the bulk compositions
 of GEMS.  They are stoichiometrically enriched in O (up to 80\%) and systematically depleted in Mg, Fe, Ca and S ($\sim$40\%) relative to solar
 abundances.   {This enigmatic abundance pattern is qualitatively similar to that of core-collapse (type II/Ibc) SN ejecta (Fig. 3).}    \citet{limongi-chieffi} have calculated the isotopic and elemental abundances in the ejecta
 of core-collapse supernovae in using a numerical hydrodynamic model and an 
 extensive nucleosynthetic network\citep{limongi-chieffi}. 
 The Limongi and Chieffi (hereafter LC03)  abundances, averaged over the Salpeter initial mass function over   13-35 M$_{\odot}$, give high O (compared to stoichiometric silicates, O/Si=2), high S compared with that inferred in typical IS grains from gas-phase depletion, and low S, Mg, Ca and Fe with respect to Si compared to solar values.  (Fig. \ref{LandC}). This is in qualitative agreement with the measurements of GEMS.

We consider chemical modifications of grains bombarded with an instantaneously
monovelocity beam of atoms with the composition of IMF-averaged SN ejecta.  Atomic or ionic
bombardment sputters target atoms from  grain surfaces and  implants projectile 
atoms into grain interiors, slowly changing their chemistry.  
\citet{gray-edmunds} have recently reexamined the question of the survival of grains during ion bombardment.  They confirm previous studies (e.g. \citet{mckee} and \citet{jones96}) that shocks in gas with cosmic abundances efficiently destroy grains.
But they also concluded that in circumstances in which the bombarding gas is very metal-rich,
implantation can dominate over sputtering  and can even lead to grain growth.       If most implanted H and He diffuse from grains and escape, but  implanted atoms  with $Z\ge6$ are retained in  grains, then implantation will dominate over sputtering if the sputtering yield, 
defined as the average number of sputtered ions per incident bombarding atom,   is less than the metallicity of bombarding gas.   In this case, bombarded grains will grow.

\begin{figure}[h!]
\begin{center}
\resizebox{!}{8cm}{\includegraphics*{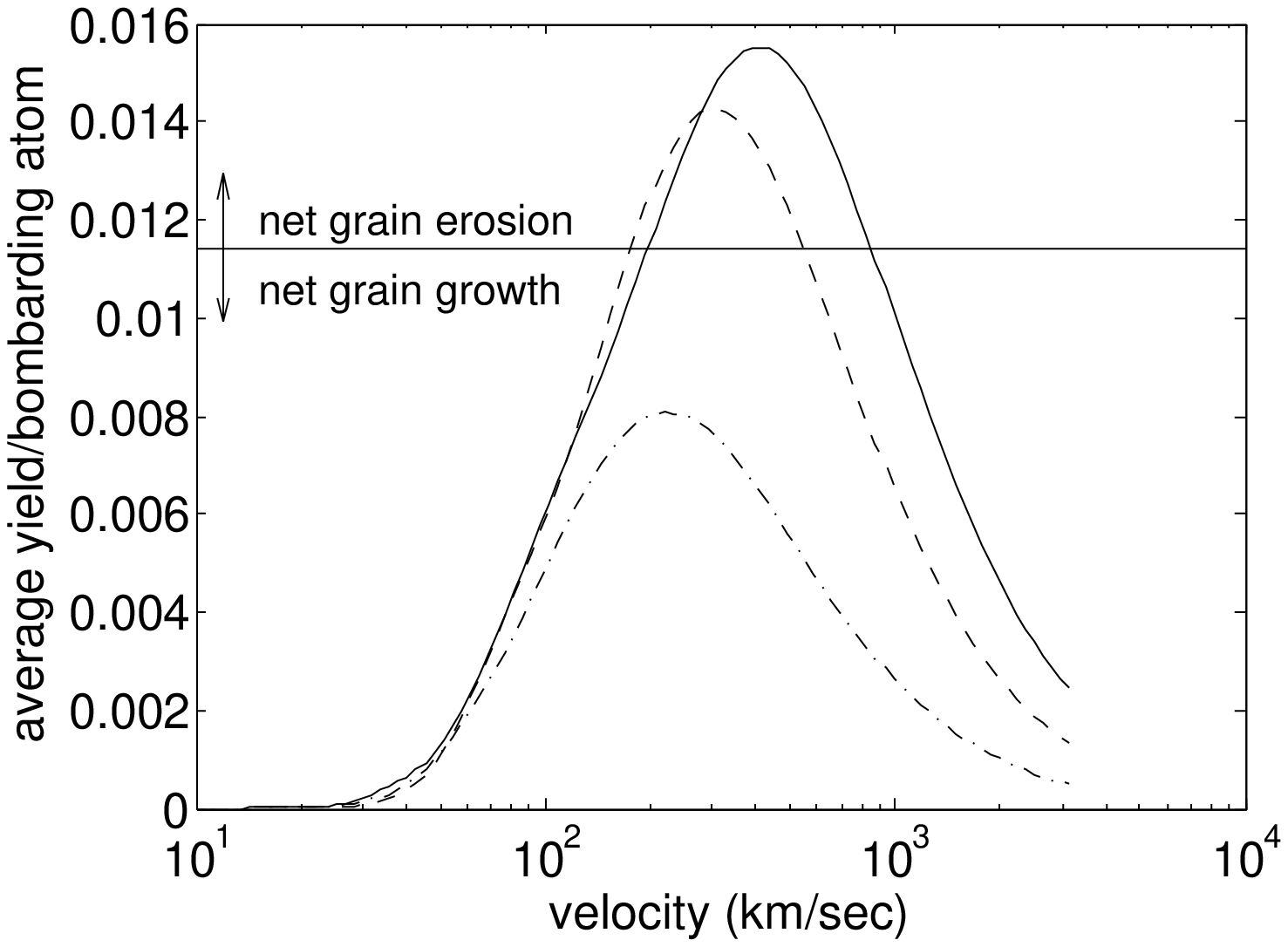}}  
\caption{Sputtering yield for FeS (solid line), SiO$_2$ (dashed line) and graphite (dash-dot line),  averaged over the Limongi and Chieffi (LC03) SN yields\protect\citep{limongi-chieffi}, as a function of velocity.   In the progression from FeS, SiO$_2$ and graphite, the peak sputtering yields occur
 at successively lower velocities (413, 308 and 218 km sec$^{-1}$ respectively).  The yields were calculated from \protect\citet{tielens}, and multiplied by 0.15.    If only H and He are lost from grains
after implanation, implantation will equal sputtering if the sputtering yield = 0.0114, the metallicity
of the  IMF-averaged LC03 SN ejecta (horizontal line).} \label{sputtering}
\end{center}
\end{figure}

Following Gray and Edmonds, we modelled
the change in chemistry  of FeS (troilite) due to intense ion bombardment with a gas with the composition of SN ejecta.  
The actual change in chemistry and morphology of the grains will be complex because of major changes in chemistry 
during bombardment.    A Monte Carlo computer code written by \citet{srim}, SRIM 2003.26,  is commonly used for sputtering
calculations at lower energies.  However, SRIM is not appropriate for the high velocities that we consider
here(Ziegler, private communication, 2004).   We calculated the sputtering yield as a function of velocity for SiO$_2$ 
bombarded by monovelocity gas with the average LC03 SN ejecta composition (Fig. \ref{sputtering}) using the analytical
 treatment of \citet{tielens}. In order of decreasing importance,  sputtering is dominated by He, H, and O.    In the absence of any experimental data, 
we assume that the sputtering of troilite is similar to that of
silicates.  In the treatment that follows, we used the expression
of Tielens {\em et al.}, but multiplied the yields by 0.15.  This value was chosen so that grains retain approximately
their original size during bombardment at 1000 km sec$^{-1}$.
We stress here that calculations of sputtering yields are subject to large uncertainties, so definitive
experimental measurements on GEMS analogs, or, preferably on GEMS themselves,
are needed to confirm (or not) their robustness to sputtering.   We discuss uncertainties in sputtering yields,
including the effect of a tough carbonaceous mantle, 
in section \ref{discussion}. 
Here we have ignored the effects of grain-grain collisions, which have the overall effect of reducing  the number of large grains with respect to smaller ones\citep{jones96}.    This may be at least partially justified because of the lower average density
of grains expected in the site that we propose in section \ref{superbubbles}.

For the purpose of illustration, 
we calculated the composition of   grains bombarded by IMF-averaged SN ejecta
 at a constant velocity of 1000 km sec$^{-1}$.  We assumed that all of the ejecta was
 in the gas phase.  We define the {\em normalized bombardment} $\xi$ to be the total mass
of bombarding atoms normalized to the original grain mass. 
 For a spherical grain of density  $\rho_{\rm g}$
and radius $r_{\rm g}$ travelling through a pathlength $x$  in
a gas of uniform density $\rho_{\rm ISM}$,

\begin{equation}
\xi  = {3\over 4} {\rho_{\rm ISM} x \over \rho_{\rm g}  r_{\rm g}}.  \label{eqn1}
\end{equation}

\begin{figure}[h!]
\begin{center}
\resizebox{!}{8cm}{\includegraphics*{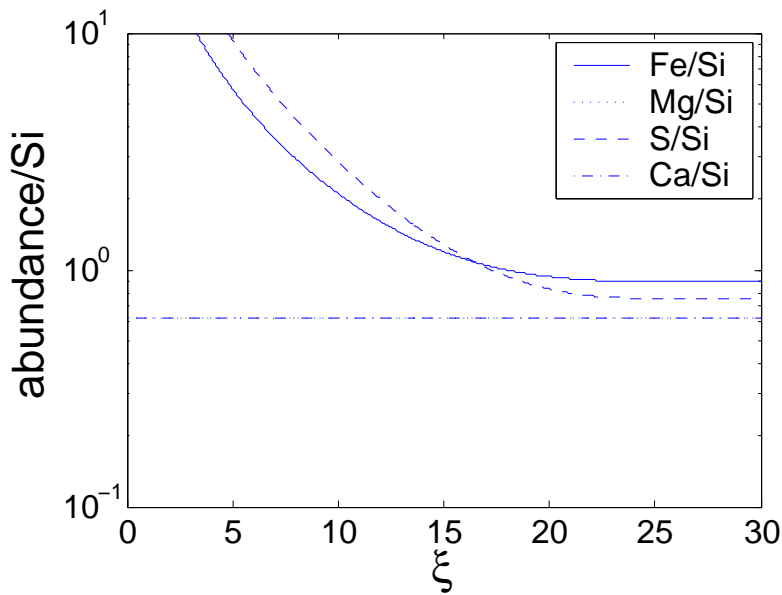}}  
\caption{Atomic ratios Fe/Si, Mg/Si, S/Si, and Ca/Si, all with respect to solar values \protect\citep{anders-grevesse} as a function of normalized bombardment $\xi$.  The bombarding atomic velocity was assumed to be 1000 km sec$^{-1}$.} \label{chemistry-xi}
\end{center}
\end{figure}

We made the simplifying assumption that the grains are completely homogenized during bombardment.  
This assumption  leads to a systematic underestimate of chemical modification: 
in reality, the grain material would be preferentially sputtered from the surface as compared with projectile atoms, which are implanted deep in the grain and are not, at least
at first, susceptible for sputtering.    We assumed that C, O, Mg, Si, and Ca 
have the same sputtering yield as Fe.  This is likely to be 
an oversimplification, but a more elaborate model is not justified given the uncertainties
in the theoretical calculation of the sputtering yield.   We assumed that H and He were immediately lost after implantation,
 but that all other elements were retained in the grain.   After bombardment by atoms of IMF-averaged SN ejecta with a normalized mass fluence 
 $\xi \gtrsim 25$, the chemistry of the grains at depths down to the maximum range of the bombarding ions 
becomes subsolar in all the elements considered, normalized to Si. (Fig. \ref{chemistry-xi}).

We next consider two astrophysical scenarios in which grains are exposed to monovelocity beams.

\section{Grain bombardment by isolated SN blastwaves}
\label{isolated-supernova}

We first consider perhaps the most obvious scenario in which dust grains are bombarded by
monovelocity SN ejecta:   exposure of dust to SN blastwaves.  
Can an encounter with a single SN blastwave achieve sufficient fluence to substantially change the
bulk composition of IS dust?   The material in the SN blastwave has two components:  the direct SN 
 ejecta and the swept-up ambient matter in surrounding ISM.    Their respective surface densities are

\begin{equation}
\sigma_{\rm ej} = {M_{\rm ej}\over 4\pi r_s^2} = 2.8\times10^{-7} {\rm g\, cm^{-2}}
\left(M_{\rm ej}\over \,15 M_\odot \right)
\left(r_s \over 30\, {\rm pc} \right)^{-2} 
\end{equation}
and 
\begin{equation}
 \sigma_{\rm swept} =  {1\over3} \rho_{\rm ISM} r_s = 7.2 \times 10^{-5}  {\rm g\, cm^{-2}}
\left(r_s \over 30\, {\rm pc} \right)
\left(A_{\rm ISM} \over 1.4\right) 
\left(n_{\rm ISM}\over 1\  {\rm cm}^{-3} \right).
\end{equation}

Now we compute the  blastwave surface density $\hat\xi$ normalized to the grain mass, that is,
the ratio of mass encountered by a grain, if it passes directly through the blastwave without deflection,
to the total mass of the grain.    This is

\begin{equation}
 \hat\xi = \left[ 
9\times10^{-3} \left(M_{\rm ej}\over 15 M_\odot  \right)
\left(r \over 30 {\rm pc} \right)^{-2} 
+
2.3 
\left(n_{\rm ism}\over 1\ {\rm cm}^{-3} \right)
\left(r \over 30 {\rm pc} \right)
\right] 
\left(a \over 100{\rm nm}\right)^{-1} 
\left(\rho_g\over 4.6{\rm g\ cm}^{-3}\right) ^{-2} 
\left(A_g \over 40\right).
\end{equation}

If grains were swept over by SN blastwaves with no deflection, their exposure factors in
a single blastwave encounter, $\xi_{\rm SN}$, would be equal to the normalized blastwave surface densities $\hat\xi$. 
We conclude that  a single encounter with a supernova remnant without deflection ($\hat\xi \sim 2$)  is insufficient to achieve the required total normalized mass fluence.    Of course, grains will
not pass straight through shocks, and should have much longer residence times than this.  
A residence time ten times longer than the straight-through transit time could result in an
exposure factor sufficient to substantially change the bulk chemistry.  But, as we discussed in
section \ref{bombardment},  grains in this environment would not  survive:  the material in the 
shocks in the average ISM will be dominated by swept-up, relatively low-metallicity material.  
Grains in such a  gas are destroyed quickly because sputtering dominates over 
implantation\citep{gray-edmunds}.

\section{Grain acceleration in SN shocks}
\label{shock-acceleration}

Ions are accelerated by supernova shocks to highly relativistic
energies.  These relativistic particles --- the galactic cosmic rays ---  fill space in the interstellar medium.
The acceleration mechanism,  diffusive shock acceleration, is almost
unbelievably efficient:  between 3\% and 10\% of the non-neutrino luminosity of supernovae
in the galaxy is channeled into the acceleration of this population of mostly relativistic particles.
The spectrum of galactic cosmic rays extends as a single power law from $\sim$ 10 GeV/amu
to $\sim10^{15}$ eV.   (Low-energy GCRs with energies less than $\sim 10$ GeV/amu are partially to completely excluded from the inner solar system by the heliosphere.
At high energies, there is a subtle but apparently sudden steeping of the spectrum at 10$^{15}$eV --- the so-called ``knee'' ---  beyond which the spectrum continues as another uninterrupted power law
to $\sim 3\times10^{18}$ eV.  Cosmic rays above this energy --- up to $3\times 10^{20}$eV --- 
are probably accelerated by a different, as yet unknown, mechanism.)

The dynamics of  charged particles in the acceleration region of the shock
are determined principally by their magnetic rigidities.  
 The maximum
energy that can be obtained by acceleration from a single SN shock is $Z \times 10^{14}$eV\citep{lagage-cesarsky}; thus, SN shocks can accelerate protons by diffusive shock acceleration
to at least 10$^{14}$eV, corresponding to a magnetic rigidity of 10$^{14}$V.   The heaviest common element, Fe,  can thus be accelerated
 to $\sim 3\times10^{15}$eV, which is close to the observed knee in the cosmic-ray all-particle spectrum.
 (The energy spectra of individual elements has not yet been directly measured at these energies.)
Building on a suggestion by \citet{epstein} that 
dust grains with the same magnetic rigidity as single ions should also be accelerated efficiently,
Ellison, Drury and Meyer have suggested that dust grains can be efficiently accelerated up to 100 keV/amu,
or approximately $0.01c$.  
The magnetic rigidity of a charged grain as a function of velocity $v$, grain average atomic number $\mu$, grain size $a$ and grain electric potential $V$ is
 
\begin{equation}
 R = 6\times 10^{12}{\rm V}
 \left(v \over 1000 {\rm \ km\ sec}^{-1}\right)
\left(\rho \over 4.6 {\rm g cm^{-3}}\right)
\left(a \over 100{\rm nm}\right)^2
\left(\phi \over 10 {\rm V}\right)^{-1}.
\label{rigidity-eqn}
\end{equation}
This is the magnetic rigidity of a $6\times10^{12}{\rm eV}$ (6 TeV)  proton.     If diffusive shock acceleration becomes
inefficient at a magnetic rigidity of $\sim10^{14}$V, as the kink in the
GCR spectrum shows must happen with charged ions, SN shocks will accelerate $a = \sim400$ nm grains  to $\sim1000 {\rm \ km\ sec}^{-1}$.     Like the galactic cosmic ray spectrum, the spectrum
of accelerated grains is expected to be approximately a power law in magnetic rigidity\citep{ellison}.
\citet{ellison} have constructed a detailed model of GCR
origin,   in which sputtered ions from fast dust grains are preferentially 
accelerated in shocks, enhancing refractory elements as compared with volatile ones.
This model for Galactic cosmic ray origin is consistent with the observed elemental composition of GCRs  \citep{MDE,westphal}.

We thus reverse the usual roles of target and projectile:
instead of being stationary and bombarded by energetic ions, grains are accelerated by SN shocks
and  are bombarded, as seen in their rest frame, by a monovelocity beam of ambient gas atoms.

\subsection{Grain slowing after acceleration}
\label{slowing}

After acceleration, grains principally slow  by collisions with atoms\citep{ellison}.
 From simple kinematics, assuming no change in cross-sectional area:
		\begin{equation}
 v(x) = v_0 \exp\left( -{\rho_{\rm ISM} A \over m} x\right)
 \end{equation}
	where $x$ is curvilinear distance, $v_0$ is the initial velocity, $A$ is the grain cross section, $m$
	is the grain mass, and $\rho_{\rm ISM}$ is the ISM density.  For a spherical grain, this is			\begin{equation}
 v(x) = v_0{e^{-x/\lambda}} 
 	\end{equation}
			where 
\begin{equation}
 \lambda = 1100\ {\rm pc} \left(\rho_g\over 4.6{\rm g\ cm}^{-3}\right) \left(n_{\rm ISM} \over 4\times10^{-3} {\rm cm}^{-3}\right)^{-1} \left(a \over 100{\rm nm}\right) 
 	\end{equation}
where $\rho_g$ is the grain density, $a$ is its diameter, and $n_{\rm ISM}$ is the ISM number density.
A grain   slows from 1000 km sec$^{-1}$ to 20 km sec$^{-1}$ in approximately four slowing lengths $\lambda$, assuming that there is no change in the cross section of the grain. 
In the frame of the grain, the fluence $\Phi$ required to slow by a factor of $e$ with respect to the gas,
assuming cosmic abundances, is 
	\begin{equation}
 \Phi = 10^{19}\, {\rm He}\  {\rm cm}^{-2} \left( a \over 100{\rm nm} \right).
 	\end{equation}
This fluence is easily sufficient  to amorphize a grain  to the depth of penetration of the ions at maximum velocity.  Even a small fraction of this slowing ($\xi \sim 10^{-3}$) can give a 
fluence sufficient to amorphize the grain rim, $\sim10^{16}$ He cm$^{-2}$ .  This fluence 
is insufficient, however, to significantly change the grain chemistry, so we conclude that, even including
post-acceleration slowing, isolated supernovae are unlikely to be responsible for GEMS synthesis.

\citet{dch} have recently proposed a model of formation of SiC
dust grains in which SiC dust particles are formed  in a dense shell  formed by a reverse shock in SN blastwaves.  An encounter with a second reverse shock then induces a large ($\sim 500$ km sec$^{-1}$) differential motion of
condensed grains and the surrounding hot metal-rich SN ejecta.
The grains are then chemically and isotopically modified by 
slowing in these ejecta in which H and He are essentially absent.  Clayton (private communication) has suggested that GEMS
may form by a similar mechanism within isolated SN blastwaves.       Although {\em prima facie}  it appears that the bulk composition (e.g., Mg/Si, Al/Si, Ca/Si) 
suggested by this model is not consistent with GEMS, it is worthwhile
to investigate whether the physical properties of GEMS --- their restricted size range, 
chemical composition and survival of relict crystals --- might be explained by a similar mechanism.

\section{Proposed site:  superbubbles blown by OB associations} 
\label{superbubbles}

Massive evolved stars generate copious amounts of dust.  
Because of their relatively short lifetimes, such massive stars are principally located near the
cloud from which they formed, and are observed astronomically 
as OB associations.   The first stellar winds in a nascent OB association
blow a quasi-spherical cavity --- a superbubble (SB) --- in the surrounding 
high-density interstellar medium\citep{mm88}.  These superbubbles are 
 filled with a hot ($>10^6$ K), tenuous ($n =1-4 \times 10^{-3}$ cm$^{-3}$) gas.  
The swept-up ambient material quickly collapses
into a thin cold shell.  Because it is in contact with the hot, tenuous interior, the cold shell
evaporates material into the SB interior.  (We consider the problem of evaporation from clouds
in section \ref{poisoning}).  The lifetime of the SB is defined by the lifetime of the longest-living supernova
progenitors, about 50 My. 

Although there is agreement that a large fraction of core-collapse (type II and type Ibc)
 supernovae occur in superbubbles, the precise fraction is not well-constrained.
 \citet{mckee} estimated that $\sim 50\%$  of core-collapse supernovae
occur in SB, while \citet{higdon}  have suggested that the fraction is closer to $\sim$90\%  
In any event, a large fraction of dust formed in circumstellar outflows in the galaxy
must pass through superbubbles before mixing with the general ISM.

The fate of dust produced by massive stars in SBs is not well understood.
We assume here that  dust  is not  swept out of the SB by SN shocks, 
but  remains near the center of the SB where it were formed.

\subsection{Composition of the ambient ISM in the superbubble}
\label{poisoning}

\begin{figure}[h!]
\begin{center}
\resizebox{!}{8cm}{\includegraphics*{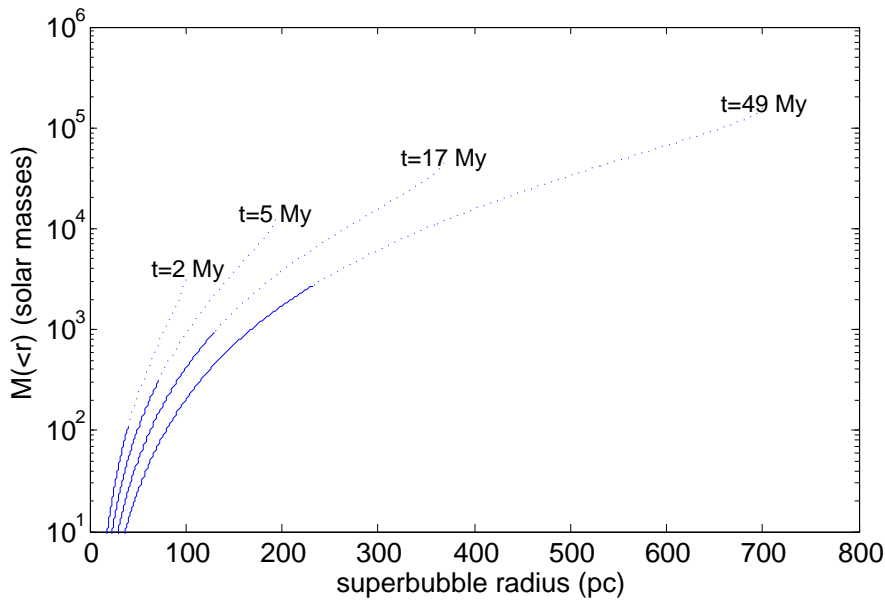}}
\caption{Integral mass inside (solid line) and outside (dotted line) a superbubble core in the spherically-symmetric Mac Low-McCray model for SB ages  2, 6, 18 and 50 My.  We assume  no mixing between the stellar ejecta in the core and evaporated material from cold shell.}
 \label{sbdensity}
\end{center}
\end{figure}

Here we adopt the idealized, spherically-symmetric superbubble model of \citet{mm88} (hereafter MM88), with an average
mechanical luminosity in supernovae of $10^{38}$ erg sec$^{-1}$, embedded in an
ISM with a density $n_0 = 1$ cm$^{-3}$. 
At an age characteristic of OB assocations\citep{mccray-kafatos} of 50 My, the SB
radius $r_{\rm SB} =700$ pc in the idealized MM88 model.  (In practice, superbubbles will 
blow out of the galactic disk before they reach this radius\citep{heiles}.) Its density profile is

\begin{equation}
n(r) = 1.4 \times 10^{-3} {\rm cm}^{-3}(1-r/r_{\rm SB})^{-2/5} .
\end{equation}
In Fig. \ref{sbdensity} we show the integral mass in a SB interior as a function of radius.  
We also adopt an average supernova rate in OB associations of 3 Myr$^{-1}$\citep{mm88}. The average number of SN in a typical OB association over its 50 My lifetime is then $\sim$150.

In the Mac Low and McCray model, the gas in the interior of the superbubble
is dominated by low-metallicity gas evaporated from the cold dense superbubble shell.  
However, this material is mostly located near the shell (Fig. \ref{sbdensity}).
The fraction of the material in the SB interior that is SN ejecta is about 1\%\citep{higdon}. 
This material will be concentrated in the core.  We assume that mixing with material
evaporated from the walls  is inefficient,  so that the interior of the SB from the center  to 0.3 $r_{\rm SB}$ (200 pc for the evolved SB)  will be highly enriched in SN ejecta.  We refer to this region of the SB interior as the metal-rich core.  

Pre-existing low-metallicity  clouds enveloped by the SB may  evaporate
material,  thereby ``poisoning''   the SB interior,   increasing its mean density and
 lowering its metallicity.  Further, fast grains that encounter clouds may stop.
\citet{mvl84} have considered the photoevaporation of clouds in HII regions.  The characteristic photoevaporation time for a 1 solar mass cloud
is of order 0.4 My, in a medium with number density 1 cm$^{-3}$ at the Str\"omgen radius,
which is of order 70pc at that number density for a single O6 star.   
(The Str\"omgen radius  in a SB  is  equal to the SB radius.) The photoevaporation time
goes only as $m^{1/2}$, so even a 100 $M_\odot$  cloud would photoevaporate in $\sim4$ My.    We conclude that clouds near the core photoevaporate early in the SB evolution, and the evaporated
gas is swept out of the SB core.  

A selection effect could operate here:  early in the evolution of the SB, clouds  poison
the interior and raise the metallicity.  Grains accelerated during this phase would be destroyed
because they would be slowing in a low-metallicity gas.  The dust that we see as GEMS has survived
because it was   accelerated later in the SB evolution, when the clouds have evaporated
and the metallicity in the core is higher.

\subsection{Propagation of shocks in the SB interior}

\citet{spitzer} gives the velocity of an adiabatic shock due to a SN with mechanical energy $E$, expanding into a medium of density $\rho  = n A_{\rm ISM} / N_A$:
\begin{equation} v_s^2 = {2 K E\over 3 \pi \rho r_s^3}  
\end{equation}
so that
\begin{equation}
 v_s = 90 {\rm km\ sec}^{-1}
\left( E \over 0.5\times 10^{51} erg\right)^{1\over2}
\left(n\over 4\times10^{-3}{\rm cm}^{-3}\right)^{-{1\over2}}
\left(r_s \over 200 {\rm pc} \right)^{-{3\over2}}.
\end{equation}
Shocks thus fall below 20 km sec$^{-1}$ in SBs at approximately 500 pc from the SB center. 
This is consistent with \citet{mm88} who found that all shocks become subsonic before they reach
the shell.  This has been confirmed experimentally by \citet{chen} who
searched for, and found no evidence of, shocked material in the shells of superbubbles in the LMC.
 
 Shocks may become radiative in the low-density SB interior, and so weaken faster as a function
of $r_s$.   Because diffusive shock acceleration is not expected to operate in radiative shocks, 
a critical question is whether shocks in the hot, low-density SB interior are primarily 
adiabatic (energy-conserving)
or radiative.   \citet{cmb88} have studied the evolution of radiative
shocks, and conclude that in a sufficiently low-density medium shocks never become radiative,
but remain adiabatic until they merge with the ambient ISM.  The critical density below which this
is true is $\sim 0.08$cm$^{-3}$ for the parameters that we have assumed for the SB interior, and with the
assumption that merging occurs at a shock speed twice the local sound speed.  
This is significantly greater than the range of densities ($1-4\times10^{-3}$cm$^{-3}$) assumed
for SB interiors, so we conclude that SN shocks in SB interiors remain adiabatic throughout their evolution.

From straightforward kinematics, assuming  a constant cross section, the normalized
bombardment for a slowing grain after a single SN encounter in the SB interior is
 $\hat\xi = \ln(1 + t_{\rm SN}/\tau_{\rm stop})$; if $t_{\rm SN}/\tau_{\rm stop} \ll 1$, then this is
 approximately:
 
 \begin{equation}
 \hat\xi =  0.3
 \left(t_{\rm SN} \over 0.3 My \right)
 \left(v \over 1000 {\rm \ km\ sec}^{-1}\right)
\left(\rho_g \over 4.6 {\rm g cm}^{-3}\right)^{-1}
\left(a \over 100{\rm nm}\right)^{-1}.
 \end{equation}

Assuming that the number of supernovae is $\sim 150$, the total normalized
bombardment $\xi$ for a 100nm grain is about 45, which is $\sim50$\% more
than is required to substantially change the chemistry of the grain; for a 500 nm grain, it is about 9.
This is not sufficient to change the grain chemistry by itself.  But this normalized bombardment
is likely to be an underestimate.  First, 
we have entirely neglected the bombardment within the shock itself; 
 this contributes $\hat\xi  = 2$ for an undeflected encounter, but since grains reside
 in the shocks until they diffuse away, this is likely to be substantially larger.
  Further, this factor assumes complete mixing of the entire grain.  This is conservative
in two ways:  first, it neglects the preserved core of the grain,  which is untouched by the bombardment;  second, as we mentioned before, it does not take into account the preferential
sputtering of grain material early in the bombardment history, which will tend to accelerate
the modification of grain chemistry as a function of normalized bombardment $\xi$.

\subsection{Grain  reacceleration in superbubbles}
\label{reacceleration}

We assume that the metal-rich SB core is swept by a supernova blastwave approximately 
every 0.3 My.   We first ask how much dust grains slow between successive blastwaves.
From straightforward kinematics, the velocity of a stopping grain is
 \begin{equation}
 v(t) = {v_0 \over 1 + t/\tau_{\rm stop}},
  \end{equation}
 where we define a characteristic stopping time 
 \begin{equation}
 \tau_{\rm stop} = {m \over \sigma_d \rho_{\rm ISM} v_0} = {\lambda \over v_0}
  \end{equation}
where $m$ is the grain mass and $\sigma_d$ is the effective cross-section for drag.  
Thus,
 \begin{equation}
 \tau_{\rm stop} = {x \over \xi v_0}
  \end{equation}
Numerically, this is 
 \begin{equation}
 \tau_{\rm stop} = 1.0 {\rm My}
\left(a \over 100{\rm nm}\right)
\left(\rho_g\over 4.6{\rm g\ cm}^{-3}\right)
\left(n_{\rm ISM}\over 4 \times10^{-3}{\rm cm}^{-3}\right)^{-1}
\left(A_{\rm ISM} \over 1.4\right)^{-1}
\left(v_0 \over 1000 {\rm \ km\ sec}^{-1}\right)^{-1}
 \end{equation}

The fact that the slowing timescale is much longer than the time between SN shock
encounters suggests that grains are continuously reaccelerated in superbubbles.
A similar idea has been suggested before in a different context.
In an effort to explain an unexpectedly large width in a measurement of  the 1.8 MeV  $^{26}$Al 
$\gamma$-ray line, \citet{sturner-naya} proposed the existence of a population of
fast ($\sim 500$ km sec$^{-1}$), freshly-synthesized grains in the ISM, maintained at high speed by
continuous reacceleration by SN shocks.   Newer measurements have indicated
that this $\gamma$-ray line is not anomalously large\citep{rhessi-al}.  Nevertheless, we revive this idea but apply it to grain
reacceleration in superbubbles.  Relocation from the warm phase of the ISM to superbubbles
 solves a problem with the original idea of Sturner
and Naya:    Galactic magnetic fields in the warm ISM prevent fast grains from propagating
 large distances, as was required in their model.  Fast grains in superbubbles do not travel
 large distances, but are swept over {\em in situ} by SN blastwaves.
   
\begin{figure}[h!]
\begin{center}
\resizebox{!}{9cm}{\includegraphics*{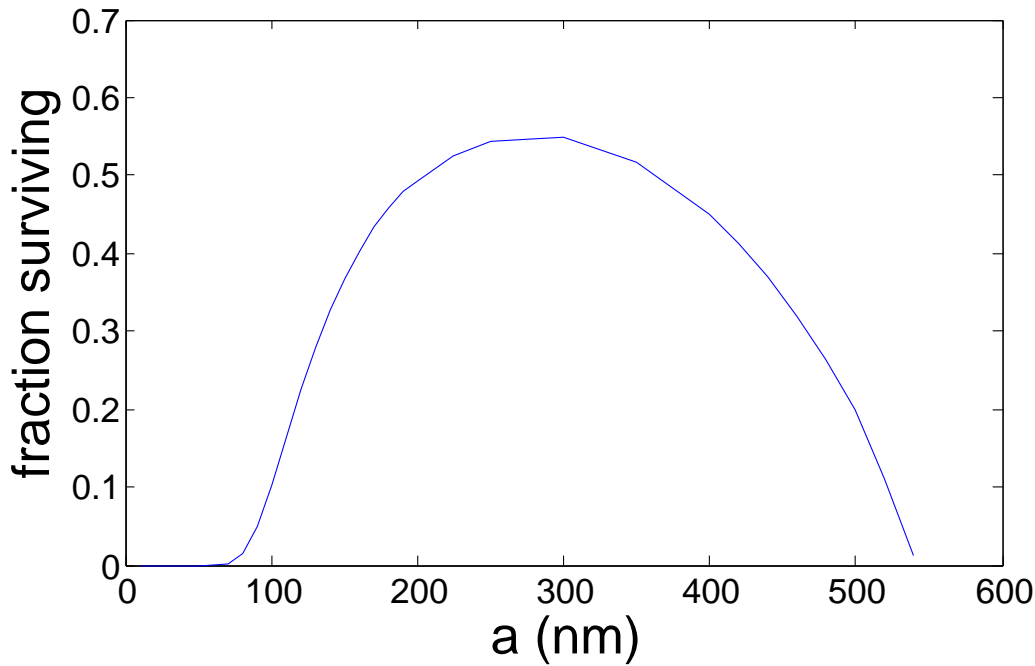}}
\caption{Fraction of grains surviving over the lifetime of a SB as a function of grain size.}
\label{survival}
\end{center}
\end{figure}

To model the behavior of reaccelerated grains in SB interiors, we used a simple Monte Carlo
 simulation of the reacceleration of 100000 individual grains in a metal-rich SB core.  
 We assumed that grains were distributed uniformly throughout the acceleration region,
 which we took to be a sphere centered on the OB association, with a radius of 150pc ---
 this is the radius from \citet{cmb88} where 
 supernova shocks merge with the ambient medium, and can no longer accelerate grains.
    We  did not attempt to model the net outward displacement that grains experience
 with each shock encounter.   We assumed  an ISM density  of $4\times10^{-3}$ cm$^{-3}$
 and a temperature $T = 10^6$ K.
  We chose the time intervals between SN shock encounters randomly
 on a Poisson distribution, with a mean interval of 0.3 My.   Each encounter was treated as follows.
 First, the grain velocity was translated from the SB rest frame to the frame of the shock.  Each grain
 then was then given a new velocity chosen randomly on a power-law distribution, with
 an index in velocity of $-2.1$, which was taken from \citet[Fig. 3]{ellison}.  We
 imposed a maximum cutoff velocity, corresponding to the observed GCR magnetic cutoff
 rigidity, $10^{14}$V, using eqn. \ref{rigidity-eqn}.  Here a major simplifying assumption was that all grains were uniformly charged to 10V.    We then translated the grain velocity back to the SB rest frame.
 We then calculated the time required for the SN blastwave to merge with the ISM from
 its current radius, and also calculated the time to the next SN encounter.  We used the minimum of those
 two times to calculate the adiabatic expansion and the collisional slowing, which were assumed
 to occur sequentially in that order.  (This is justified since the timescale for adiabatic expansion
 is much shorter than the slowing timescale.)  The adiabatic expansion was treated as:
  \begin{equation}
 v_{\rm exp} = v_0 \left[r \over \min(r_{\rm merge},r_{++})\right]^{{3\over2}(\gamma-1)} 
  \end{equation}
where $v_0$ and $v_{\rm exp}$ are the velocities before and after expansion,  $r_{\rm merge}$ is the radius of the SN blastwave when it merges with the ISM, $r_{++}$ is the radius of the SN blastwave
 at the time of the {\em next} SN shock encounter, and $\gamma=5/3$.
We assumed that grains that experience velocities at or below 400 km sec$^{-1}$, near
the peak of sputtering yield, are sputtered away and do not survive.  In Fig. \ref{survival} we show the
fraction of grains surviving over the SB lifetime as a function of grain size.  

\begin{figure}[h!]
\begin{center}
\resizebox{!}{12cm}{\includegraphics*{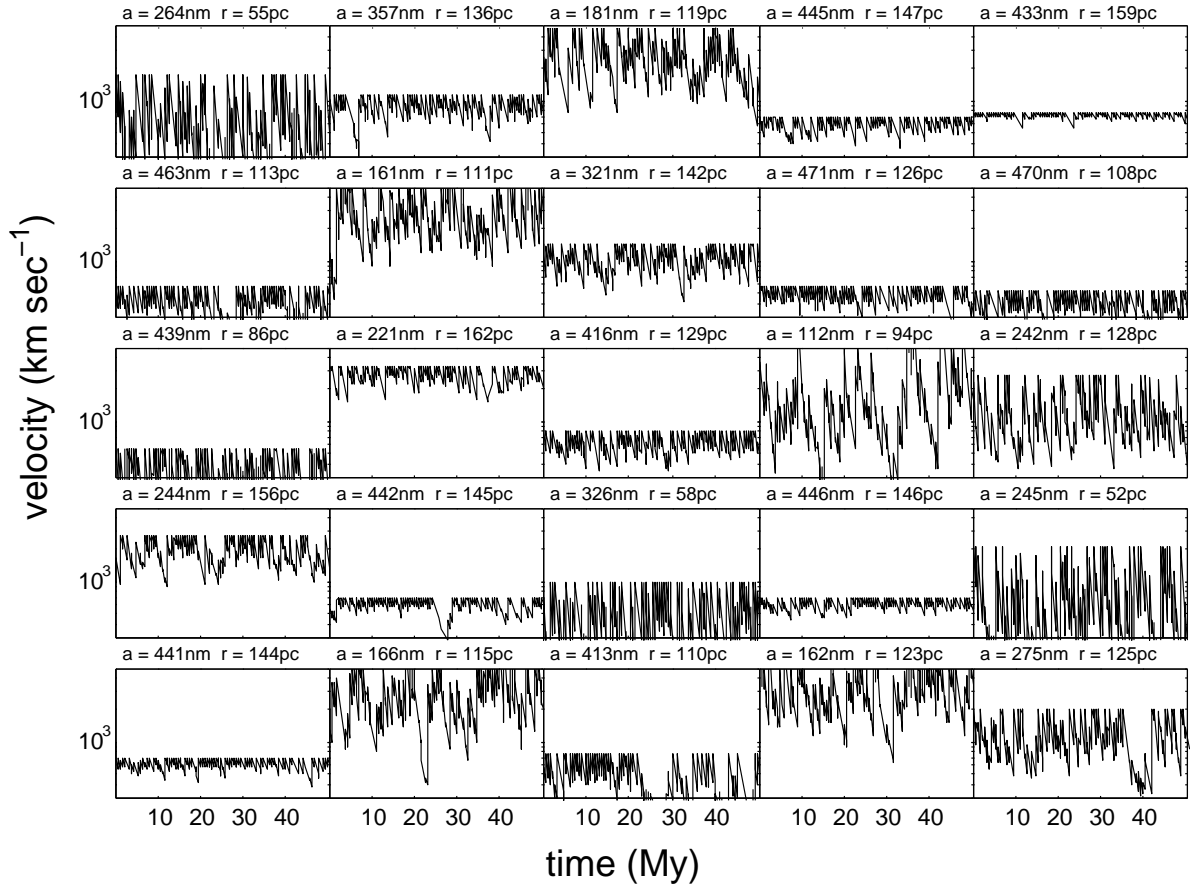}}
\caption{Individual velocity histories for 25  troilite grains, uniformly distributed in size between
$a = 100$ and $a=500$nm, and uniformly distributed in the SB core.  Grain sizes and radial
distances from the SB core center are shown above each plot.}\label{histories}
\end{center}
\end{figure}

 In Fig. \ref{histories} we show the individual velocity histories for 25 surviving grains. 
 Grains near the center of SB core experience severe adiabatic losses, and are systematically slower
 than those closer to the SB core boundary.

\begin{figure}[h!]
\begin{center}
\resizebox{!}{9cm}{\includegraphics*{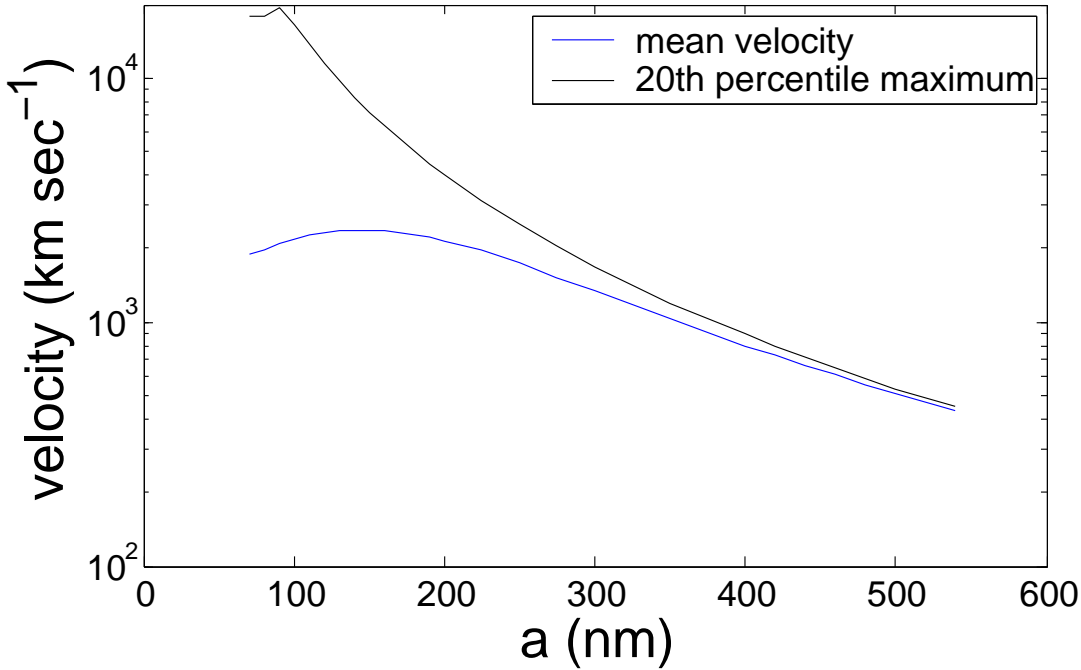}}
\caption{Average velocity (bottom curve) of surviving grains and their 20th-percentile maximum velocity, $v_{80}$,  the velocity  that is exceeded by 80\% of grains at some
point in their individual histories (top curve).} \label{velocities}
\end{center}
\end{figure}

 The boundary of the relict { crystal} for an individual grain is determined by the range
of the {\em fastest} velocity that that particular grain has achieved.  
For some grains,  the fastest velocities are not much above the average, but  in many cases, very high velocities, several thousand km sec$^{-1}$, 
are at least temporarily achieved --- these grains would be completely amorphized, and would
contain no relict { crystals}.    Approximately 20\% of grains contain relict { crystals}; in our model, then, 80\% of grains are achieve sufficiently large velocities that they are completely amorphized.
In order to check that  this fraction is  consistent with our model, we followed
100000 reaccelerated grains of fixed size, propagating in the same environment described above.
In Fig. \ref{velocities}, we show the average velocity $\bar{v}$ and $v_{80}$,  the velocity
that is exceeded by 80\% of the surviving grains.   At $a=350$ nm,  $v_{80} \sim1100$ km sec$^{-1}$.
No studies have yet been done of the distribution of GEMS rim thicknesses, but the observed range in
one particular 350nm GEMS grain ($\sim$150 nm, Fig. 1) gives a velocity, $\sim900$ km sec$^{-1}$, that
 is remarkably similar to $v_{80}$ for this size.

\subsection{Confinement in the superbubble core}

An important question is whether the accelerated grains are effectively confined to metal-rich
core of the superbubble.  If they travelled in straight lines at 1000 km sec$^{-1}$ ($\sim $1000 pc Myr$^{-1}$) , they could easily escape from the superbubble core.  The grains are charged, however, so will
 diffuse through the ambient magnetic field inside the SB.   This magnetic field
is essentially completely unknown.  We 
make the assumption that, at worst,  the SB interior magnetic field scales with the gas density from the galactic ISM, 
so that it is at least 10 nG inside the SB.  

The curvilinear distance $d_{\rm SN}$ travelled between shock encounters is of order
$d_{\rm SN} \sim v_0 t_{\rm SN} = 300$ pc.  From \citet{ellison}, the Larmor radius for a charged grain is $r_l = R/Bc$ is
\begin{equation}
 r_l = 0.16 {\rm pc}
 \left(v \over 1000 {\rm \ km\ sec}^{-1}\right)
\left(\mu \over 25\right)
\left(a \over 100{\rm nm}\right)^2
\left(\phi \over 10 {\rm V}\right)^{-1}
\left(B \over 10{\rm nG}\right)^{-1}.
\end{equation}
We assume that  the scattering length is of order the Larmor radius.    Since $\lambda \ll d_{SN}$,  this is essentially a diffusion problem:   over the SB lifetime the grains diffuse a linear distance
  \begin{equation}
d_{\rm diff} = \sqrt{r_l d_{SN}} 
  \sim 85 pc  \left(v \over 1000 {\rm \ km\ sec}^{-1}\right)
\left(\mu \over 25\right)^{1/2}
\left(a \over 100{\rm nm}\right)
\left(\phi \over 10 {\rm V}\right)^{-1/2}
\left(B \over 10 {\rm nG}\right)^{-1/2}.
\end{equation}
Thus, small grains are effectively confined  to  metal-rich core ($\sim 300$pc radius) by the ambient magnetic fields if the ambient magnetic field is at least 10 nG.
With this value of the field, 500nm grains would just be able to diffuse out of the 
near the end of the SB lifetime.    We discuss this assumption in the section \ref{discussion}.
We point out that GCRs are {\em not} similarly confined, since, at a given magnetic rigidity,
they have a velocity that is $\sim 300$  times larger than grains.

\section{Do GEMS constitute the long-sought GCR source?}
\label{connection}

As we mentioned earlier in this paper,  the enhancement in abundance in GCRs of refractory elements (e.g., Fe, Pt) as compared with volatiles (e.g., S, He, Ne, Ar, Pb)
may be most easily understood if  GCR nuclei originate in shock-accelerated dust grains\citep{MDE, ellison}.  
The Pb/Pt ratio is particularly diagnostic in this regard\citep{westphal}.  It is natural to ask
 how the GCR elemental composition compares with that of the  IMF-averaged LC03 SN ejecta.
\citet{engelmann} have reported GCR source abundances as derived from measurements  by HEAO-3-C2.  Their elemental ratios at the GCR source, normalized to the local galactic (equivalent to solar system) values of \citet{anders-grevesse}, are:  
  [C/Si] = 0.415 $\pm$ 0.047,
   [O/Si] =  0.219 $\pm$ 0.022,
   [Mg/Si] =  0.97 $\pm$ 0.072,
   [S/Si] =  0.267 $\pm$ 0.040,
   [Ca/Si] =  0.954 $\pm$ 0.16,
  [Fe/Si] =  0.900 $\pm$ 0.265.
  
The S and O abundances at the GCR source are depressed with respect to Si,
 in qualitative accord with the observed S and O abundance in GEMS.  (C is also depressed with respect
 to Si, but C has never been measured in GEMS, principally due to contamination problems.)    However, Mg, Ca and Fe are apparently solar at the GCR source, within 
 the reported error bars, which range from $\sim$7\% for Mg to $\sim$30\% for Fe. 
 Thus, the comparison does not appear at first to be encouraging, although
  the  error bars for the source compositions, which are due principally to uncertainties in
   propagation corrections,
  are still consistent with substantially depressed abundances for these elements
  as compared with solar values.

\citet{LRZ}
have suggested that GCRs  originate specifically in dust grains condensed from SN ejecta,
and point out that the GCR source composition compares well with that of SN ejecta, {\em if
type Ia supernovae are included.}    In their model,  dust grains that condense in the supernova
ejecta overtake the SN shock, making their refractory elements available for sputtering and 
injection into the acceleration region.   Since their original paper, precision measurements of GCR 
isotopic composition by the Advanced Composition Explorer (ACE)\citep{ni59} have
shown  that the electron-capture isotope $^{59}$Ni is almost completely absent in GCRs.
Because $^{59}$Ni is a clock that measures the time between nucleosynthesis and acceleration,
\citet{ni59} were able to use its nonobservation to set a lower
 limit on that time of $\sim10^5$ y, ruling out the possibility that fast dust overtakes the shock.
 However, the $^{59}$Ni result still allows the possibility that supernova shocks
 can accelerate dust, and therefore cosmic rays, from the ejecta of previous generations
 of supernovae in superbubbles, since the average time between SN shocks in superbubbles
 is much longer than the decay time of $^{59}$Ni.
 
Although they may originate from a common source, GCRs and GEMS can still have
quite different bulk compositions,  because GEMS are highly selected.    GCRs are accelerated
 ions sputtered from grains in shocks of both  core-collapse supernovae, which predominantly
 occur  in superbubbles, and accretion-induced collapse (type Ia) supernovae, which
 occur essentially uniformly in the ISM.  As we have described, 
 fast dust accelerated by II/Ibc supernovae can survive sputtering and implantation in
 the metal-rich superbubble core; but regardless of whether a dust grain survives or not, ions
 sputtered from it  will contribute to the GCRs.   The grains that
 survive in this harsh environment are observed in IDPs as GEMS, 
 but grains smaller than $\sim 100$nm, which
 constitute a large fraction of the dust mass, do not survive and so are not represented in IDPs.   
 Dust accelerated by most
 type Ia supernovae is expected to be old dust in the ISM:  this dust on average will have
  a solar composition.    However, because dust accelerated in this environment will
  slow principally in low-metallicity gas, it is efficiently destroyed and so is not represented
  in IDPs.    Thus, our model does not require that bulk GCR and GEMS compositions be similar.
  However, our picture does suggest that GEMS are a surviving population of 
  accelerated dust grains that are the source of GCR nuclei.
 
 { The majority of GEMS have isotopic compositions that are solar within measurement errors.  
  The isotopic composition of GCRs has been now well-measured for all major elements
  through Ni\citep{wiedenbeck-rev}.  Before precise measurements 
  were available, it was widely assumed that GCR isotopic abundances would diverge strongly
  from solar values, so it was a surprise when GCRs were found to be essentially solar
  in isotopic composition, with the exception of a strong excess of $^{22}$Ne and
  a mild excess of $^{58}$Fe\citep{wiedenbeck-rev}.  Although excesses in 
  $^{22}$Ne and $^{58}$Fe could be due to a small admixture of material from Wolf-Rayet stars,
  additional anomalies (e.g., in Mg isotopes) should be seen that are not observed.  No consistent
  explanation for these excesses has yet been proposed.  Nevertheless, similar
  excesses should be present in GEMS if these anomalies are due to GCRs originating in superbubbles.
Some GEMS do exhibit isotopic anomalies.  This may be simply residual isotopically anomalous
 circumstellar material, or in may be due to exposure to isotopically anomalous SN blastwaves.}

 \section{Discussion}
 \label{discussion}
 
 \subsection{Restricted size range of GEMS}

  In Fig. \ref{survival}, we have shown in our model 
  that only grains in the size range $100-500$ nm survive
  in the SB interior.    In this model, we neglected the possibility of complete
  penetration of grains by high-velocity atoms, which will further accelerate the destruction of small
  grains.   In grains that are large compared to the range of bombarding ions, implantation
can dominate over sputtering since bombarding atoms stop inside the grain, and
 sputtering occurs only on entry.  Such grains can grow with time.  In contrast,
 grains that are small compared to the range of bombarding ions are quickly eroded because bombarding atoms do not stop inside the grain, and
 sputtering occurs both on entry and on exit, so that sputtering dominates over implantation.
 Further, since the ions leaving the grain are slower, they will generally be more effective at sputtering (Fig. \ref{sputtering}), so the effective sputtering rate will be more than twice that of a large grain.  We have confirmed this effect using SRIM.

\begin{figure}[h!]
\begin{center}
\resizebox{!}{10cm}{\includegraphics*{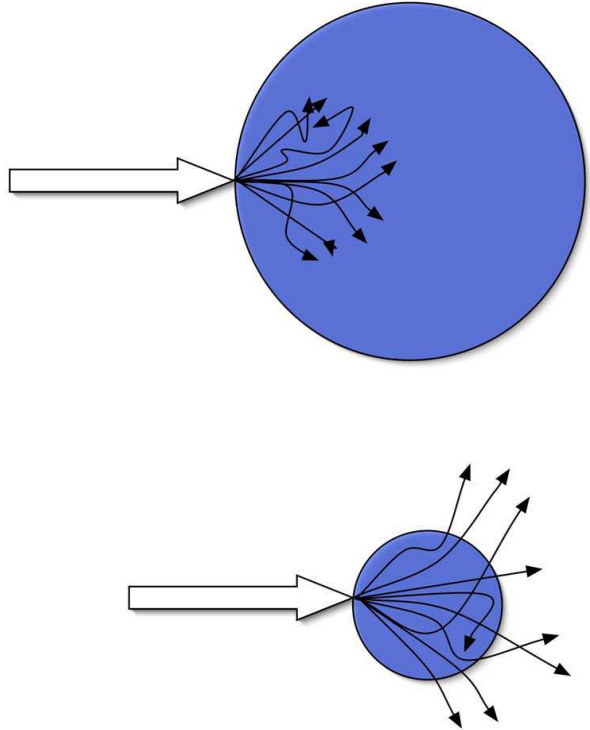}}
\caption{For grains that are large compared to the range of bombarding ions (top), implantation
can dominate over sputtering since bombarding atoms stop inside the grain, and
 sputtering occurs only on entry.  Such grains can grow with time.  In contrast,
 grains that are small compared to the range of bombarding ions (bottom) can be quickly eroded because bombarding atoms do not stop inside the grain, and
 sputtering occurs both on entry and on exit, so that sputtering dominates over implantation.} \label{small-sputtering}
\end{center}
\end{figure}

\subsection{Are GEMS typical IS grains?}

Astronomical observations have shown that the vast majority 
of interstellar silicates are amorphous \citep{kemper}.  The fact that GEMS are also amorphous
together with the similarity between the 10$\mu$m feature of IDPs and that of interstellar dust 
   has led to the suggestion that GEMS are typical representatives of the interstellar dust \citep{bradley94, bradley99}.
The fact that some GEMS exhibit isotopic anomalies indicative of a circumstellar origin support this idea.
 \citet{keller-messenger}  have recently pointed out, however,  that the bulk chemistry of GEMS is inconsistent with the typical elemental composition of interstellar dust as derived from gas phase depletions\citep{gas-depletions}.   They propose that GEMS formed by multiple mechanisms in multiple environments (in the solar system, in  presolar environments, and in the interstellar medium, for example). Based on the bulk compositions of GEMS-rich IDPs, they conclude that most
  GEMS formed in the solar system.     However, the bulk compositions of GEMS and the IDPs that contain them are equally inconsistent with a solar system origin \citep{schramm, bradley-ireland,bradley94}.  Moreover, the presence of relict { crystals} and the restricted size range of GEMS are not easily understood in terms of a solar system origin.   Our model is also consistent with the observation that while some GEMS have non-solar O isotopic compositions most have normal (solar) compositions (Keller and Messenger, 2004).  The model predicts that most of the O atoms in most GEMS were deposited from the gas phase leading inevitably to a highly averaged (solar) O isotopic composition.  Nevertheless, nothing in the model that we propose
 requires that GEMS be representative of average interstellar dust.   
 The absence of average interstellar dust  --- amorphous silicates with bulk chemistry consistent with ISM gas phase depletions --- in IDPs is unexplained.   This paper implies that it is appropriate to revisit the topic of abundances of the major solid-forming elements, particularly Si, as derived from gas-phase depletions in the ISM. 

We have proposed a mechanism by which GEMS may be formed from shock-accelerated
fast dust in superbubbles, and which explains for the first time the basic properties of 
of GEMS,  their restricted size range, their compositions, and the presence of surviving internal relict { crystals} in some GEMS.
Our model is tied intimately to  the acceleration of galactic cosmic
rays;   indeed, parameters in our model are determined from direct observations of galactic cosmic
rays in the solar system.
The agreement between the predicted and observed GEMS rim thicknesses
 is particularly striking when we consider that this model includes only one unconstrained parameter ---
 the sputtering correction factor for SRIM.
  All other parameters are fixed either theoretically (e.g., the size of the SB core) or
 experimentally (e.g., the maximum cutoff rigidity is fixed by the observed GCR energy spectrum).

We have made a number of assumptions in our model, which we list here:

$\bullet$  We made the {\em ad hoc} assumption that sputtering yields are substantially smaller 
($\sim{1\over6}$) than the values predicted for silicates by \citet{tielens}.  
Although the uncertainties in sputtering yields are subject to uncertainties of a factor of two,
we require a more substantial suppression of sputtering rates for GEMS.     It is expected that many
interstellar grains acquire a thin carbonaceous mantle\citep{tielens} that should
suppress sputtering yields substantially.  Indeed, the non-thermal sputtering yield of graphite due to He
bombardment for velocities $> 700$ km sec$^{-1}$ is less than 40\% of the yield of silicates
\citet[Fig. 11]{tielens}.  If such a carbonaceous layer does indeed protect GEMS
from sputtering, it must be thin if the external shape of relict { crystals} is to reflect that of the glassy
rims.   If this is the case, it may lead to an instability in sputtering:  reaccelerated grains that
slow to velocities near the peak of the sputtering yield may first lose their protective carbonaceous
mantle, then are rapidly eroded because of the enhanced sputtering rate of the bare silicate surface.
The preponderance of { relict crystalline FeS} ($\sim$20\% of GEMS) remains to be explained, as does the  comparative rarity of relict silicates.  We speculate that silicate minerals may sputter
more readily than FeS.

$\bullet$  We assumed that the magnetic field in the SB interior scales, at worst, with the ambient
gas density, so is no smaller than 10 nG.   In fact, the measured GCR spectrum strongly suggests
   that  the magnetic field is orders of magnitude larger than this.  There is a long-standing problem
in GCR astrophysics:  isolated SN shocks should not be capable
of accelerating particles much beyond 10$^{14}$eV, and yet 
 the GCR spectrum extends continuously, without steps, from 100 MeV through the knee
 to the ankle at $\sim 3\times10^{18}$ eV.
The continuity and smoothness of the spectrum strongly suggests that 
all GCRs over this energy range must be accelerated by the same mechanism.  
\citet{bykov-toptygin}  have recently suggested that GCRs below the 
knee at $10^{15}$ eV have been accelerated by individual SN, but GCRs above the knee, 
up to the ``ankle'' in the GCR spectrum at $\sim 3\times10^{18}$ eV, have been accelerated by multiple
SN encounters.  Acceleration through the collaborative effects of multiple SN in an association is a compelling picture, but it requires
a magnetic field of 30$\mu$G in the SB interior --- more than four orders of magnitude
larger than the minimum field strength required to confine fast grains to the metal-rich core.   

$\bullet$  We   assumed that the metal-rich SB core near the OB association mixes 
inefficiently with material evaporated from the cold SB shell, and that cloud 
poisoning is negligible.  The extent of mixing is not well-constrained
observationally. 

$\bullet$  Finally, and crucially, we assumed that the $10^6$K gas in the metal-rich SB core has the same composition as the  IMF-averaged LC02 core-collapse SN ejecta, that is, that the amount
of material depleted by grain condensation is negligible.  Dust condenses quickly
in rapidly-cooling SN ejecta, as observed in SN 1987A.  However, we assume that 
even highly refractory elements from dust destroyed in the hot SB interior does not recondense,
and that {\em most}  dust in SN ejecta is destroyed.   This assumption
is consistent with our picture, since small ($<100$ nm) grains, which contain most of the dust mass, 
are readily destroyed in the SB interior.

Our model makes specific predictions for future observations.  First, 
$^{22}$Ne/$^{20}$Ne should be much larger ($\sim 5\times$) 
than the solar value in GEMS, just as it is in GCRs.
We would expect systematic
compositional differences between grains originating as pyrrhotite as compared with the more rare
grains that originate in other types, because of the presence of residual material from
the original grain.  For example, GEMS containing pyrrhotite relict
{ crystals} should have larger bulk S than those containing forsterite or enstatite.
The decay products of  $^{26}$Al and $^{60}$Fe may  be present, but may not be
detectable, unless there is substantial inhomogeneity in Mg/Al or  Fe/Ni ratio in the SB interior
that would allow the fossil radioactivities to be detected through positive correlations
between, e.g., $^{26}$Mg/$^{24}$Mg  and $^{27}$Al/$^{24}$Mg.
Similarly, although there is no evidence of $^{59}$Ni in the ``bulk''  GCRs, 
it is possible that relict $^{59}$Ni could be found in GEMS if there is substantial 
inhomogeneity in Ni/Co in the SB interior. 
Our model also predicts that GEMS   exhibit
a $^{58}$Fe excess if the GCR $^{58}$Fe excess originates from core-collapse (type II/Ibc) SN ejecta in superbubbles.
So far as we know, no study has yet been made of  the  distribution of amorphous rim thicknesses as
a function of grain size.  Our model predicts that relict { crystals} will preferentially be found in
large ($>200$nm) grains.

\bibliographystyle{unsrt}

\begin{thebibliography}{99}
\bibitem[Bradley(1994)]{bradley94} Bradley, J. P. 1994,  { Science} { 265}, 925
\bibitem[Bradley et al.(1999)]{bradley99} Bradley, J. P., et al. 1999,  { Science} { 285}, 1716
\bibitem[Bradley \& Dai (2004)]{bradley-dai04} Bradley, J. P. and Dai, Z. R. 2004,  ApJ, in press.
\bibitem[Bradley \& Ireland(1996)]{bradley-ireland} Bradley, J. P.  \& Ireland, T.Ê 1996, ``The search for interstellar components in interplanetary dust particles".Ê In Physics, Chemistry, and Dynamics of Interplanetary Dust (APS Conference Series, Vol 104, 1996) Bo A. S. Gustafson \& M. S Hanner (eds.).
\bibitem[Bykov \& Toptygin(2001)]{bykov-toptygin} Bykov, A. M \& Toptygin, I. N. 2001, { Astron. Lett} { 27} 625 trans. from 2001, {Pis'ma v astronomicheskij zhurnal} { 27}, 735
\bibitem[Cardelli(1994)]{gas-depletions} Cardelli, J. 1994, { Science} { 265}, 209
\bibitem[Carrez et al.(2002)]{carrez} Carrez, P.{\em et al.} 2002, {MAPS} { 37}, 1599
\bibitem[Chen et al.(2000)]{chen} Chen, C.-H. { \em et al.} 2000, { ApJ}, { 119}, 1317
\bibitem[Chioffi, McKee \& Bertschinger(1988)]{cmb88} Cioffi, D., McKee, C. F., \& Bertschinger, E. 1988, { ApJ} { 334}, 252
\bibitem[Demyk et al.(2001)]{demyk} Demyk, K. {\em et al} 2001, A\&A 368, L38
\bibitem[Deneault, Clayton \& Heger (2003)]{dch} Deneault, E. A.-N., Clayton, D. D., \& Heger, A. 2003 ApJ 594, 312
\bibitem[Engelmann et al.(1990)]{engelmann} Engelmann, J.J. {\em et al.} 1990, { A\&A} { 233}, 96
\bibitem[Ellison, Drury \& Meyer(1997)]{ellison} Ellison, D. C.,  Drury, L. O'C., \& Meyer, J.-P. 1997,  { ApJ} { 487} 197
\bibitem[Epstein(1980)]{epstein} Epstein, R. I. 1980, { MNRAS} { 193} 723
\bibitem[Gaisser(1990)]{gaisser} Gaisser, T. K. 1990,  ``Cosmic Rays and Particle Physics,'' Cambridge Univ. Press 
\bibitem[Gray \& Edmunds(2004)]{gray-edmunds} Gray, M. D. \& Edmunds, M. G. 2004, MNRAS{  349}, 491
\bibitem[Anders \& Grevesse(1988)]{anders-grevesse} Grevesse, N. \& Anders, E. 1988,  AIP Conf. Proc. { 188}, 1
\bibitem[Higdon, Lingenfelter \& Ramaty(1998)]{higdon} Higdon, J. C., Lingenfelter, R. E. \& Ramaty, R. 1998,{ ApJ} { 509}, L33
\bibitem[Jones(1996)]{jones96} Jones, A. P., Tielens, A. G. G. M \& D. J. Hollenbach. 1996, ApJ 469, 740 
\bibitem[Jones(2000)]{jones2000} Jones, A. P. 2000, JGR A5, 10257
\bibitem[Keller \& Messenger(2004)]{keller-messenger} Keller, L. P. \& Messenger, S. 2004,  { Proc. Lunar Planet. Sci. Conf.} abs.  \#1985
\bibitem[Kemper \& Tielens(2003)]{kemper} Kemper, F. \& Tielens, A. G. G. M.  2003,  Proc. Conf. { Astrophysics of Dust} ed. A. N. Witt 
\bibitem[Koo, Heiles \& Reach(1992)]{heiles} Koo, B.-C., Heiles, C., \& Reach, W. T. 1992, { ApJ} { 390}, 108
\bibitem[Lagage \& Cesarsky(1983)]{lagage-cesarsky} Lagage, P. O. \& Cesarsky, C. J. 1983, { A\&A} { 125} 249.
\bibitem[Limongi \& Chieffi(2003)]{limongi-chieffi} Limongi, M. \& Chieffi, A. 2003, { ApJ} { 592}, {404} 
\bibitem[Lingenfelter, Ramaty \& Kozlovsky(1998)]{LRZ}  Lingenfelter, R. E., Ramaty, R., \& Kozlovsky, B. 1998, { ApJ} { 500}, L53
\bibitem[McCray \& Kafatos(1987)]{mccray-kafatos} McCray, R. \& Kafatos, M.  1987, { ApJ} { 317}, 190
\bibitem[Mac Low \& McCray(1998)]{mm88} Mac Low, M.-M. \& McCray, R. 1988, { ApJ} { 324}, 776
\bibitem[McKee, Van Buren \& Lazareff(1984)]{mvl84} McKee, C. F., Van Buren, D. \& Lazareff, B. 1984,  { ApJ} { 278}, L115
\bibitem[McKee(1989)]{mckee}  McKee, C. F. 1989, IAU Series 135, 431
\bibitem[Meyer, Drury \& Ellison(1997)]{MDE} Meyer, J.-P., Drury, L. O'C., \& Ellison, D. C. 1997, { ApJ} { 487} 182
\bibitem[Schramm et al.(1999)]{schramm} Schramm {\em et al.} 1999, { Meteoritics} 24, 99
\bibitem[Spitzer(1978)]{spitzer} Spitzer, L. 1978, ``Physical Processes in the Interstellar Medium,'' Wiley \& Sons
\bibitem[Sturner \& Naya(1999)]{sturner-naya} Sturner, S. J. \& Naya, J. E.. 1999, { ApJ} { 526}, 200
\bibitem[Smith(2003)]{rhessi-al}  Smith, D. M. 2003, ApJL 589, L55
\bibitem[Tielens et al.(1994)]{tielens} Tielens, A. G. G. M., McKee, C. F., Seab, C. G., \& Hollenbach, D. J. 1994, { ApJ} { 431}, 321
\bibitem[Westphal et al.(1998)]{westphal} Westphal, A. J. {\em et al.} 1998,  { Nature} { 396}, 50 
\bibitem[Ziegler \& Biersack(2003)]{srim} Ziegler, J.  F. \& Biersack, J. P., SRIM code (www.srim.org)
\bibitem[Wiedenbeck et al.(1999)]{ni59} Wiedenbeck, M. { et al.} 1999,  { ApJ}, { 523}, L61
\bibitem[Wiedenbeck et al.(2001)]{wiedenbeck-rev} Wiedenbeck, M. E. { et al.}. 2001, { Spac. Sci. Rev.} { 99}, 15
\end{thebibliography}

\typeout{References here}

\noindent\textbf{\large Acknowledgments}\\

We are deeply indebted to Don Ellison and Chris McKee for their thoughtful comments and
advice in preparing this manuscript.  We thank Don Clayton for criticism and suggestions.
We also thank Lindsay Keller, Steven Sturner and James Ziegler for helpful conversations.   AJW was supported by NASA grant NNG04GI27G;  JPB was supported by NASA grants NAG5-10632 and NAG5-10696.
\ \\
 
 Correspondence and requests for materials should be addressed to AJW. (e-mail: westphal@ssl.berkeley.edu).

\end{document}